\documentclass[11pt]{article}
cm \voffset = -2truecm

\usepackage{amssymb}

\newcommand{\bra}{\begin{array}}
\newcommand{\era}{\end{array}}
\newcommand{\beq}{\begin{equation}}
\newcommand{\eeq}{\end{equation}}
\newcommand{\beqar}{\begin{eqnarray}}
\newcommand{\eeqar}{\end{eqnarray}}

\def\BC{\bb C}
\def\_\BC{\bbi C}

\def\( {\left(}
   \def\) {\right)}
\def\[ {\left[}
\def\] {\right]}
\def\no2 {{\textstyle{n\over 2}}}

\def\dag {{\dagger}}

\newcommand{\om}{\omega}

\newcommand{\lam}{\lambda}
\newcommand{\si}{\sigma}

\newcommand{\eps}{\epsilon}

\newcommand{\te}{\theta}

\newcommand{\pa}{\partial}
\newcommand{\al}{\alpha}

\newcommand{\lga}{\longrightarrow}

\newcommand{\da}{\dagger}

\newcommand{\sq}{\sqrt}

\newcommand{\lb}{\label}

\newcommand{\PR}[1]{ {\it Phys.~Rev.} {\bf #1}}
\newcommand{\PRL}[1]{ {\it Phys.~Rev.~Lett.} {\bf #1}}

\begin{document}
\begin{titlepage}
\setcounter{page}{1}
\renewcommand{\thefootnote}{\fnsymbol{footnote}}

\begin{flushright}
ucd-tpg:08.02\\
arXiv:0806.3574
\end{flushright}

\vspace{5mm}
\begin{center}

{\Large \bf 
 Periodic Structures with Rashba Interaction\\ in
Magnetic Field}

\vspace{5mm}

{\bf Ahmed Jellal$^{a,b,}$\footnote{ajellal@ictp.it,
jellal@ucd.ac.ma}} and {\bf Rachid
Hou\c{c}a$^{b,}$\footnote{rachid.houca@gmail.com}}

\vspace{5mm}

{\em $^{a}$The Abdus Salam International Centre for Theoretical Physics},\\
{\em  Strada Costiera 11, 34014 Trieste, Italy}\\


{ \em $^{b}$Theoretical Physics Group,  
Faculty of Sciences, Choua\"ib Doukkali University},\\
{\em Ibn Ma\^achou Road, PO Box 20, 24000 El Jadida,
Morocco}\\

\vspace{30mm}

\begin{abstract}

We analyze the behaviour of a system of particles living on a
periodic crystal in the presence of a magnetic field $B$. This can
be done by involving a periodic potential $U(x)$ and the Rashba
interaction of coupling constant $k_{so}$. By resorting the
corresponding spectrum, we explicitly determine the band structures
and the Bloch spinors. These allow us to discuss the system
symmetries in terms of the polarizations where they are shown to be
broken. The dynamical spin will be studied by calculating different
quantities. In the limits: $k_{so}$ and $U(x)=0$, we analyze again
the system by deriving different results. Considering the strong $B$
case, we obtain an interesting result that is the conservation of
the polarizations. Analyzing the critical point
$\lam_{k,\si}=\pm\sq{1\over 2}$, we show that the Hilbert space
associated to the spectrum in $z$-direction has a zero mode energy
similar to that of massless Dirac fermions in graphene. Finally, we
give the resulting energy spectrum when $B=0$ and $U(x)$ is
arbitrary.

\end{abstract}
\end{center}
\end{titlepage}




\section{Introduction}

Unlike the quantum Hall effect (QHE)~\cite{prange} that can be
resulted from the impurities~\cite{klitzing} or the Coulomb
intercation~\cite{tsui}, the spin Hall effect (SHE)~\cite{perel} is
essentially due to something else. In fact, this phenomena can be
seen as consequence
 of the  Rashba or Dresselhaus interactions~\cite{kato,wunderlich,rashba}. They
 depend what kind of spin-orbit coupling
is involved and more precisely they are originated from two
different effects. For the first one is caused by the asymmetry of
the confining quantum well but the second is a bulk
effect~\cite{feve}. The spin-orbit coupling
 leads to a spin accumulation in
the direction transverse to the applied electric field. These show
the importance of such interaction and motivate their analysis in
different contexts.


Many theories appeared as good candidates to describe different
issues related to the Rashba interaction. Different questions have
been addressed and lead to interesting developments on the
subject. In particular 
an analytical and numerical results for the effect of such
spin-orbit coupling on band structure, transport, and interaction
effects in quantum wires when the spin precession length is
comparable to the wire width have been reported in~\cite{zulike}.
The Harper--Hofstadter problem for two-dimensional electron gas with
Rashba interaction subject to periodic potential and perpendicular
magnetic field has been studied analytically and
numerically~\cite{perov}. One may cite also other interesting papers
related to the subject such
as~\cite{peeter,demikhovskii,khomitsky,plyushchay}.

Very recently, an analysis has been appeared dealing with other
issues. This concerns the investigations of the basic features of a
two-dimensional electron gas with Rashba interaction and two
in-plane potentials superimposed along directions perpendicular to
each other~\cite{smirnov}. The first of these potentials is assumed
to be a general periodic potential while the second one is totally
arbitrary. A general form for Bloch's amplitude is found and an
eigenvalue problem for the band structure of the system is derived.
Moreover the general result is applied to the two particular cases
in which either the second potential represents a harmonic in-plane
confinement or it is zero. It is found that for a harmonic
confinement regions of the Brillouin zone with high polarizations
are associated with the ones of large group velocity.

Motivated by the investigations of different groups cited above and
in particular~\cite{smirnov}, we develop our proposal to deal with
different issues. For this, we consider a system of particles living
on the plane $(x,z)$ in the presence of an external magnetic field.
To do our task, we include a period potential $U(x)$ as well as the
Rashba interaction. By splitting the corresponding Hamiltonian, we
diagonalize different parts to end up with the total eigenvalues and
Bloch spinors for the two first transverse modes, i.e $n=0,1$. These
will be used to show that there is a hidden symmetry in the
spectrum.

On the other hand, by analyzing the polarizations we conclude that
such symmetries are not preserved for the present system. Moreover,
we study the dynamical spin by evaluating the velocity and spin
current operator, which are used to give the corresponding Hall
conductivity. Moreover, we discuss two limits where the first is
when the Rashba coupling constant and this leads to end up with the
Landau problem submitted to the periodicity constraint in
$x$-direction. The second one is $U(x)=0$ that allows us to get the
same Landau problem but amended with the Rashba interaction.

To deeply investigate the basic features of the present system, we
treat an interesting case. Indeed, for a strong $B$ we analyze the
corresponding spectrum and symmetries to end up with different
conclusions. As an interesting consequence, we show that the
polarization symmetries are preserved in such case and this
generalize the obtained result in~\cite{smirnov} to the magnetic
field. More importantly, such polarizations are obtained to be
$B$-dependent and therefore it can be controlled by different values
taken by the magnetic field.

On the other hand, we consider the critical point
$\lam_{k,\si}=\pm\sqrt{1\over 2}$ that appears as a singularity of
the spectrum to underline its influence on the obtained results.
More precisely, by resorting different quantities we notice the
spectrum corresponding to $z$-direction is sharing some commun
features with the graphene. In particular, a zero mode energy will
come up in similar way to massless Dirac fermions. In the end, we
analyze the periodic structures when $B$ is switched off and by
considering two limits we recover some known results like the Rashba
dispersion relation for instance.

The present paper is organized as follows. In section $2$, for later
convenience we study the Bloch theory for one-dimensional system and
establish the Hamiltonian for two-dimensional where the Rashba
interaction is included. This will be generalized by introducing a
magnetic field $B$ perpendicular to the periodic crystal in section
$3$ in order to create a confinement in one-direction. Moreover, we
split the corresponding Hamiltonian into two parts to get an easy
diagonalization. In section $4$, we treat different parts to obtain
the eigenvalues and eigenstates and for the total spectrum we
restrict ourselves to the two first modes. In section $5$, we
consider two different limits: $k_{so}=0$ and  $U(x)=0$ to
characterize the system behaviour in such cases. We analyze the
polarization symmetries by evaluating different components in
section $6$ and show that only that in $z$-direction is broken where
the remaining are nulls. We discuss the spin dynamical properties by
evaluating some relevant quantities in section $7$. We consider the
case when $B$ is strong to deduce different conclusions in section
$8$. The critical point will be analyzed in section $9$. We consider
the periodic structure of the system when $B$ is switched of in
section $9$. Finally, we conclude our work.

\section{Bloch theory for one-dimensional system}

We start by showing the spinorial structure influence on
one-dimensional periodic crystal. This can be done by making use of
a Bloch theory and characterize the system behaviour. In doing so,
let us consider a simple Hamiltonian, which is given by
\beq\label{h1d0} H_{0}^{\sf 1D}={{\hbar}^2k_x^2\over2m}+U(x) \eeq
where  $U(x)$ is a periodic potential, such as \beq
U\left(x+L\right)=U\left(x\right) \eeq and $L$ is its period along
the $x$-direction. Clearly, without  $U(x)$ we simply have a plane
wave in one-dimension as solution of the system.

Note that, $H_{0}^{\sf 1D}$ is a simple problem that can easily be
solved. Indeed, the eigenvalues $\varepsilon_{l,\si}^{(0)}(k_B)$ and
eigenstates $|l,k_B,\si\rangle$ are solutions of the equation \beq
H_{0}^{\sf
1D}|l,k_B,\si\rangle=\varepsilon_{l,\si}^{(0)}(k_B)|l,k_B,\si\rangle.
\eeq They are labeled by the Bloch's quasi-momentum $k_B$ that is a
discrete value in the Brillouin zone and the band index $l$.  The
energies $\varepsilon_{l,\si}^{(0)}(k_B)$ are degenerates with
respect to the spin index because one can obtain the relations \beq
\varepsilon_{l,+1}^{(0)}(k_B)=\varepsilon_{l,-1}^{(0)}(k_B)\equiv
\varepsilon_{l}^{(0)}(k_B) \eeq where $\varepsilon_{l}^{(0)}(k_B)$
read as
\begin{equation}\label{blen}
\varepsilon_{l}^{(0)}(k_B)={\hbar^2 k_B^2\over 2m}.
\end{equation}
To get the corresponding Bloch wavefunctions,  we simply project the
states $|l,k_B,\si\rangle$ on the coordinate representation
$\{|x,\si\rangle\}$. Doing this process to get \beq \langle
x,\sigma'|l,k_B,\sigma\rangle={1\over\sqrt{L_0}}e^{ik_Bx}u_{l,k_B,\sigma}(x,\sigma')
\eeq where  $u_{l,k_B,\sigma}(x,\sigma')$ is the Bloch amplitude and
$L_0$ is the system size. Moreover, we can easily establish the
relation \beq
u_{l,k_B,\sigma}(x,\sigma')=\delta_{\sigma',\sigma}u_{l,k_B}(x).
\eeq Note that,  $u_{l,k_B}(x)$ is also  a periodic function and
satisfies \beq u_{l,k_B}(x)=u_{l,k_B}(x+L). \eeq In conclusion, the
spinorial structure of the Bloch theory is trivial in the present
case. This suggests to see what happens if some interaction in terms
of spin are included. This issue has been addressed for a system in
two-dimensions and submitted to a confining
potential~\cite{smirnov}. This work will be generalized to get
others interesting results and offer different discussions.

Before proceding our generalization by including the magnetic field
$B$, it is relevant to fix the starting point that is $B=0$. This
will allow us to underline the difference between these two cases
and therefore make some comments. In fact,
 a Bloch theory in two-dimensional
periodic crystal with  in the spin-orbit coupling described by the
Hamiltonian \beq\lb{ham} H_0={{\hbar}^2 \vec{k}^2\over2m}-{{\hbar}^2
k_{so}\over m} \left(\si_{x}k_z-\si_{z}k_x\right)+U(x)+V(z) \eeq was
considered in the reference~\cite{smirnov} where the second term is
the Rashba interaction and the confining potential is fixed to be
\beq\label{cpot} V(z)={m\om_0^2\over 2}z^2. \eeq Some interesting
results have been derived and in particular the polarization
symmetries have been discussed. In fact, we will show that these
results can be recovered as a particular case of the present
proposal when $B$ is strong and then it can be used to controle the
polarization behaviours.

\section{Gauge field coupling}

To deal with different issues we introduce an external magnetic
field $B$ to end up with a confining potential in similar way to
(\ref{cpot}) but with a frequency $B$-dependent. This will allow us
to obtain an appropriate Hamiltonian that will be used to do our
task and in particular to generalize the result obtained
in~\cite{smirnov} to the case of $B$ and offer some comments.

Let us start by considering a periodic crystal parametrized by
$(x,z)$ in the presence of a perpendicular $B$. The Hamiltonian for
a single particle can be written as \beq\lb{1}
H={{\vec{\pi}}^2\over2m}-{{\hbar}^2 k_{so}\over
m}\left(\si_{x}{\pi}_z-\si_{z}{\pi}_x\right)+U(x) \eeq where the
conjugate momentum is \beq \vec{\pi}=\vec{p}-{e\over c}\vec{A}. \eeq
Unlike (\ref{ham}) the Rashba term is actually $B$-dependent and
there is no external confinement analogue to the potential $V(z)$.
However, we will see that
  an analogue form to $V(z)$ can spontaneously be created thanks
to the emergence of $B$. Thus, one can expect that this effect will
make difference with respect to the results obtained
in~\cite{smirnov}. The corresponding quasi-momentum becomes \beq
\vec{K}=\vec{k}-{e\over \hbar c}\vec{A} \eeq where $\vec{A}$ is the
vector potential. These materials will be the starting point to
establish a tool in order to tackle different problems in this work.

In this level, it is necessary to fix some quantities involved in
the above Hamiltonian. In doing so, we choose the Landau gauge
\beq\label{landga} \vec{A}=B(z,0). \eeq This allows (\ref{1}) to
take another form, such as
\begin{equation}\label{totham}
H= {1\over 2m}\left[p_z^2 + \left(p_x-{eB\over c}z \right)^2 \right]
-{\hbar^2 k_{so}\over m} \left[\si_x p_z -\si_z \left(p_x-{eB\over
c}z \right)\right] +U(x).
\end{equation}
One can use different methods to solve this problem. In our
analysis, we propose to split $H$ into two parts and each one will
be treated separately. They are \beq H=H^{(1)}+H^{(2)} \eeq where
the Hamiltonian $H^{(1)}$ is given by \beq\lb{H'}
H^{(1)}={\hbar^2\over 2m}\left(k_x+{\sigma}_zk_{so}\right)^2+U(x)
-{\hbar^2k_{so}^2\over 2m}+{\hbar^2k_z^2\over 2m}+{1\over
2m}\left({eB\over
  c}\right)^2z^2-{\hbar eB\over mc}\left(k_x+\si_z
 k_{so}\right)z
\eeq and $H^{(2)}$ is fixed to be \beq
 H^{(2)}=-{\hbar^2k_{so}\over m}\si_xk_z.
\eeq Clearly, due to the presence of $B$, the Bloch spinors will be
changed from $|l,k_B,\si\rangle$ to the states $|l,k_B,\eta\rangle$.
In fact, they verify the eigenvalue equation
\begin{equation}\label{hamtot}
H|l,k_B,\eta\rangle=\varepsilon_{l,\eta} (k_B)|l,k_B,\eta\rangle
\end{equation}
where the parameter $\eta$ will be fixed later. Then, the next step
is to solve such equation in order to derive the whole spectrum that
can be used to deal with different issues and in particular discuss
the symmetries.

\section{Eigenvalue problems}

As it will be clear soon, the above splitting is useful is sense
that the whole spectrum can be obtained in the easiest way. More
precisely, we resort different spectrum entering in the game by an
immediate derivation from the Bloch theory
 and making an algebraic approach to overcome
some difficulties. In doing so, we split $H^{(1)}$ into two forms as
well and analyze each part to get its spectrum. As far as the whole
spectrum is concerned, we restrict ourselves to two first modes
those allow us to make comparison with the results obtained
in~\cite{smirnov}. On the other hand, this restriction is argued by
dealing with the Rashba interaction effect on the band
structures~\cite{zulike}.

\subsection{$H^{(1)}$ spectrum}

By separating the Hamiltonian $H^{(1)}$ into two parts, we show that
the first spectrum can be easily obtained from that corresponding to
the zero magnetic field case seen before. However, the second one
can be deduced from one-dimensional harmonic oscillator. This can be
done by distinguishing two different cases: spin down and up. But,
it is enough to handle only one of them and other can be obtained in
similar way to reach the conclusion.

The simplest way to derive the corresponding spectrum is to divide
$H^{(1)}$ itself into parts. The advantage of this is to refer to
each part independently and also to make diffrent comparisons with
other cases. Then, the first one reads as
\begin{equation}
H^{(1)}_x={\hbar^2\over 2m}\left(k_x+{\sigma}_zk_{so}\right)^2+U(x)-
{\hbar^2k_{so}^2\over 2m}
\end{equation}
and the second is given by
\begin{equation}\label{hamzd}
H^{(1)}_z= {\hbar^2k_z^2\over 2m}+{1\over 2m}\left({eB\over
  c}\right)^2z^2-{\hbar \om_c}\left(k_x+\si_z k_{so}\right)z
\end{equation}
where $\om_c={eB\over mc}$ is the cyclotron frequency. It is now
clear that the second term in (\ref{hamzd}) plies the role of the
confining potential (\ref{cpot}). This will allow us to make contact
with one-dimensional harmonic oscillator in order to get the
spectrum of (\ref{hamzd}).

The $H^{(1)}_x$ form is easy to handle where its
 eigenvalues and eigenstates can be derived
in similar way to those obtained in section $2$. In fact, one can
find the spectrum \beq
\varepsilon_{l,\si}^{(1)}(k_B)|_x=\varepsilon_{l}^{(0)}
\left(k_B+\si k_{so}\right)- {\hbar^2k_{so}^2\over 2m}
\eeq where $\si=\pm 1$ is the spin index. It can also be written as
\begin{equation}\label{hxsp}
\varepsilon_{l,\si}^{(1)}(k_B)|_x={\hbar^2\over 2m} \left(k_B^2+2\si
k_{so}\right).
\end{equation}
Clearly in the absence of the Rashba coupling, i.e. $k_{so}=0$, we
recover  $\varepsilon_{l}^{(0)}(k_B)$ that corresponds to spinless
particle.

In deriving the  $H^{(1)}_z$ spectrum, we use some changes to end up
with the harmonic oscillator form that is easy to diagonalize. We
proceed by distinguishing two cases: spin up and down, i.e. $\si=\pm
1$, where we treat only one of them and the second will follow
immediately. Then, we start by defining new variables for $\si=1$,
such as \beq P={p_z\over \sqrt{2m\hbar\om_c}}, \qquad
Q=\sqrt{{m\om_c\over 2\hbar}}z. \eeq They verify the commutation
relation \beq \left[Q,P\right]={i\over 2}. \eeq In terms of $P$ and
$Q$, $H^{(1)}_z|_{\si=1}$ takes the form \beq
H_z^{(1)}|_{\si=1}=\hbar\om_c\left[P^2+Q^2-\sqrt{{2\hbar\over
m\om_c}} \left(k_B+k_{so}\right)Q\right] \eeq which is sharing some
commun features with the harmonic oscillator. To clarify this point,
let us make an algebraic analysis by introducing the annihilation
and creation operators. They can be realized as
\begin{equation}
a=Q+iP+\lambda_{k,+1}, \qquad a^{\dagger}=Q-iP+\lambda_{k,+1}
\end{equation}
where the quantity $\lambda_{k,+1}$ is a function of the Bloch and
rashba vectors, such as \beq
 \lambda_{k,+1}=-\sqrt{{\hbar\over 2m\om_c}}\left(k_B+ k_{so}\right).
\eeq It is easy to check the relation \beq \left[a,
a^{\dagger}\right]=\mathbb{I}. \eeq With these operators,
$H^{(1)}_z|_{\si=1}$ can be mapped as \beq\label{mapham}
H^{(1)}_z|_{\si=1}=\hbar\om_c\left[a^{\dagger}a+{1\over2}(1-2\lambda_{k,+1}^2)\right].
\eeq To make contact with the harmonic oscillator spectrum,
firstly we define a new frequency in terms of $\om_c$ and
$\lambda_{k,+1}$. This is \beq\lb{om}
\om_{+1}=\om_c\left(1-2\lambda_{k,+1}^2\right) \eeq which allows us
to write (\ref{mapham}) as \beq
H^{(1)}_z|_{\si=1}=\hbar\om_{+1}\left({a^{\dagger}a\over
1-2\lambda_{k,+1}^2}+{1\over 2}\right). \eeq Note that, hereafter we
assume that the condition $\lambda_{k,+1}\neq\pm \sq{1\over 2}$ is
fulfilled. In fact, we return to analyze this critical point
separately and underline its influences on the system behaviour.
Secondly, we can consider new operators in terms of $a$ and $a^\da$
involved before. They are \beq b={1\over \sq
{1-2\lambda_{k,+1}^2}}a, \qquad b^{\dagger}={1\over
\sq{1-2\lambda_{k,+1}^2} }a^{\dagger}. \eeq They lead to the
Hamiltonian \beq
H_z^{(1)}|_{\si=1}=\hbar\om_{+1}\left(b^{\dagger}b+{1\over
2}\right). \eeq This is nothing but the harmonic oscillator
Hamiltonian and therefore the spectrum can easily be obtained.
Indeed, the eigenvalues are given by \beq\lb{13}
\varepsilon_{l,n,+1}^{(1)}(k_B)|_z= \hbar\om_{+1}\left(n+{1\over
2}\right), \qquad n=0,1,2\cdots \eeq and the eigenstates read as
\beq\lb{13p} |n\rangle={(b^\da)^n\over \sq{n!}}|0\rangle. \eeq

To complete the above analysis we need to consider the second case
where $\si=-1$. Indeed, the same calculations remain valid and the
only thing that should be changed is $\lambda_{k,+1}$. Its analogue
can be defined in similar way, such as \beq
\lambda_{k,-1}=-\sqrt{{\hbar\over 2m\om_c}}\left(k_B- k_{so}\right).
\eeq It is easy to observe $\lambda_{k,+1}$ and  $\lambda_{k,-1}$
verify the symmetry relation
\begin{equation}\label{lamsr}
\lambda_{k,+1} (k_B)= -\lambda_{k,-1} (-k_B).
\end{equation}
Therefore, from the above analysis we end up with the eigenvalues
for spin down case, such as
\begin{equation}\label{-1sp}
\varepsilon_{l,n,-1}^{(1)}(k_B)|_z= \hbar\om_{-1}\left(n+{1\over
2}\right), \qquad n=0,1,2\cdots
\end{equation}
where $\om_{+1}$ and $\om_{-1}$ satisfy a similar relation to
(\ref{lamsr}). These last spectrum's will be used to analyze the
critical point. This will allow to  make contact with the Dirac
fermions in the presence of magnetic field.

At this level, we have all ingredients to get the $H^{(1)}$
spectrum. Indeed, we just combine all together to obtain those
verify the eigenvalue equation \beq
H^{(1)}|l,k_B,n,\si\rangle=\varepsilon_{l,\si}^{(1)}(k_B)|l,k_B,n,\si\rangle
\eeq which is showing that the eigenvalues of $H^{(1)}$ are of the
form \beq
\varepsilon_{l,n,\si}^{(1)}(k_B)=\varepsilon_{l}^{(0)}\left(k_B+\si
k_{so}\right)- {\hbar^2k_{so}^2\over
2m}+\hbar\om_{\si}\left(n+{1\over 2}\right). \eeq It is clear that
the corresponding eigenstates can be obtained as tensor product,
namely
\begin{equation}\label{h1st}
|l,k_B,n,\si\rangle = |l,k_B,\si\rangle \otimes |n\rangle.
\end{equation}
The $H^{(1)}$ eigenfunctions can be obtained by projecting the above
states on the coordinate representation. Thus, they can be found in
terms of the Bloch amplitude as \beq \langle
x,n',\si'|l,k_B,n,\si\rangle={1\over
  \sqrt{L_0}} \delta_{n',n}\delta_{\si',\si}e^{ik_Bx}u_{l,k_B+\si k_{so}}(x).
\eeq
 With this we complete the derivation of the eigenvalues and the
eigenstates problems of $H^{(1)}$. These will be used to derive
other quantities and discuss different issues.

\subsection{Total spectrum}

After getting the $H^{(1)}$ spectrum we need to determine that
corresponding to  $H^{(2)}$ in order to end up with the total
eigenvalues and eigenstates. In this case, the Bloch amplitude
acquires a new spinoral structure, such as
\begin{equation}
\langle x,n,\si|l,k_B,\eta\rangle={1\over
 \sqrt{L_0}} e^{ik_Bx}u_{l,k_B,\eta}(x,n,\si)
\end{equation}
where  $u_{l,k_B,\eta}(x,n,\si)$ is also a periodic function, namely
\begin{equation}
u_{l,k_B,\eta}(x,n,\si) = u_{l,k_B,\eta}(x+L,n,\si).
\end{equation}
Before proceeding, let us denote by
$\te_{l,k_B,\eta}\left(n,\si\right)$
 the Bloch spinors in the $\{l,k_B,n,\si\}$ representation,
where $\eta$ will be fixed later. They can be written as \beq
\te_{l,k_B,\eta}\left(n,\si\right)=\langle
l,k_B,n,\si|l,k_B,\eta\rangle \eeq which allowing us to derive the
relation \beq \langle l',k_B',n,\si|l,k_B,\eta\rangle=
\delta_{l',l}\delta_{k_B',k_B}\te_{l,k_B,\eta}\left(n,\si\right).
\eeq

To explicitly determine $\te_{l,k_B,\eta}\left(n,\si\right)$, we
have to solve the eigenvalue equation (\ref{hamtot}) where still
only $H^{(2)}$ to be treated. In doing so, let us project this
equation on the states $|l,k_B,n,\si \rangle$  to get the Bloch
spinors as
\begin{eqnarray}\lb{5}
&&\sum_{n',\si'}\left[\langle
l',k_B',n,\si|H^{(1)}|l,k_B,n',\si'\rangle +\langle
l',k_B',n,\si|H^{(2)}|l,k_B,n',\si'\rangle\right]
\te_{l,k_B,\eta}\left(n',\si'\right)= \nonumber\\
&& =
 \varepsilon_{l,\eta(k_B)}\delta_{l',l}\delta_{k_B',k_B}
\te_{l,k_B,\eta}\left(n,\si\right).
\end{eqnarray}
This is actually involving two kinds of matrix elements where each
one can separately be evaluated because they are related to
$H^{(1)}$ and $H^{(2)}$. Indeed, for the first it is easy to obtain
\beq\lb{4} \langle l',k_B',n,\si|H^{(1)}|l,k_B,n',\si'\rangle
=\varepsilon_{l,n,\si}^{(1)}(k_B)\delta_{l',l}\delta_{k_B',k_B}\delta_{\si,\si'}.
\eeq The second kind can be written as \beq\label{sprt} \langle
l',k_B',n,\si|H^{(2)}|l,k_B,n',\si'\rangle =-{\hbar^2k_{so}\over
m}\langle\si|\si_{x}|\si'\rangle\langle n|k_z|n'\rangle
\delta_{l',l}\delta_{k_B',k_B}. \eeq This contains also two
separated matrix elements. As concerning the first one, its is easy
to end up with the result \beq
\langle\si|\si_{x}|\si'\rangle=1-\delta_{\si,\si'}. \eeq Combining
all together to write the final equation for the eigenvalues
$\varepsilon_{l,\eta}(k_B)$ and eigenspinors
$\te_{l,k_B,\eta}(n,\si)$ as
\begin{eqnarray}\label{fequa}
&&\sum_{n',\si'}\left[
\varepsilon_{l,n,\si}^{(1)}(k_B)\delta_{l',l}\delta_{k_B',k_B}\delta_{\si,\si'}
-{\hbar^2k_{so}\over m} \left(1-\delta_{\si,\si'}\right) \langle
n|k_z|n'\rangle
\delta_{l',l}\delta_{k_B',k_B}\right]\te_{l,k_B,\eta}\left(n',\si'\right)=\nonumber\\
&& = \varepsilon_{l,\eta(k_B)}\delta_{l',l}
\delta_{k_B',k_B}\te_{l,k_B,\eta}\left(n,\si\right).
\end{eqnarray}
Actually, this is depending to different quantum numbers and can be
solved by making use of some restrictions. This will be the subject
of the remaining subsection.

To solve the above equation, we may focus on some cases those allow
us to simplify different quantities. Indeed, assuming that we have
the same band indices $l'=l$ and Bloch's quasi-momentum $k_B=k_B'$,
then (\ref{fequa}) can be reduced to
\begin{equation}\lb{6}
\sum_{n',\si'}\bigg\{\delta_{n,n'}\delta_{\si,\si'}
\varepsilon_{l,n,\si}^{(1)}(k_B) -{\hbar^2k_{so}\over
2m}\left(1-\delta_{\si,\si'}\right)\langle
  n|k_z|n'\rangle \bigg\}\te_{l,k_B,\eta}(n',\si') =
\varepsilon_{l,\eta}(k_B)\te_{l,k_B,\eta}(n,\si).
\end{equation}
To go further, we should evaluate the second term in (\ref{6}). This
can be done by returning to the annihilation and creation operators
in order to obtain \beq A-A^{\dagger}={2i\hbar\over
\sq{1-2\lambda_{k,\si}^2}}P. \eeq This is leading to write $k_z$ as
\beq k_z=i\left(A-A^{\dagger}\right)
\sq{{m\om_c\over2\hbar}\left({1\over2}-\lambda_{k,\si}^2\right)}.
\eeq Now it is easy to get the matrix element of $k_z$ as
\beq\label{mtelt} \langle n|k_z|n'\rangle=\pm i\delta_{n,n'\pm1}
\sqrt{{m\om_{\si}\over 2\hbar}{\left(n+{1\over 2}\mp{1\over
2}\right)}}. \eeq Therefore, the eigenvalue equation takes the form
\begin{eqnarray}\lb{bfequa}
&&\sum_{n',\si'}\left\{\delta_{n,n'}\delta_{\si,\si'}
\left[{\hbar^2\over 2m}(k_B^2+2\si k_{so})+
\hbar\om_{\si}\left(n+{1\over 2}\right)\right]
\mp i {\hbar^2k_{so}\over 2m}\left(1-\delta_{\si,\si'}\right)
\delta_{n,n'\pm1} \sqrt{{m\om_{\si}\over 2\hbar} {\left(n+{1\over
2}\mp{1\over 2}\right)}} \right\}
 \nonumber\\
&&
\times \te_{l,k_B,\eta}(n',\si')
=\varepsilon_{l,\eta}(k_B)\te_{l,k_B,\eta}(n,\si).
\end{eqnarray}
Many discussions can be reported on this spectrum as well as its
corresponding eigenfunctions. In the next, we treat a special case
that allows us to make contact with other results and give different
comments. More precisely, since we are wondering to make comparison
with the proposal developed in~\cite{smirnov}, we restrict ourselves
only on the two first transverse modes, i.e. $n=0,1$. On the other
hand, such approximation has been motivated by analyzing the spin
accumulation in quantum wires with strong Rashba
coupling~\cite{zulike}. This restriction will allow us to simply
(\ref{bfequa}) and derive interesting results.

Focussing on $n=0,1$ and $\si=\pm 1$, (\ref{bfequa}) reduces to a
diagonalization of $4\times 4$ matrix, which is a task that can
analytically be done. This process leads to fix $\eta$ as a number
running on $(1,\cdots,4)$ and to end up with two relations between
different energies. They are
\begin{equation}\lb{7}
\varepsilon_{l,\eta=1,2}(k_B)=\varepsilon_l^+(k_B)-\xi_{l_{1,2}}(k_B),\qquad
\varepsilon_{l,\eta=3,4}(k_B)=\varepsilon_l^+(k_B)+\xi_{l_{2,1}}(k_B)
\end{equation}
where $\varepsilon_l^+(k_B)$ reads as
\begin{eqnarray}
\varepsilon_l^+(k_B) = {1\over 2}
\left[\varepsilon_l^{(0)}(k_B+k_{so})+
\varepsilon_l^{(0)}(k_B-k_{so})\right]+
\hbar\om_c\left(1-2\lambda_{k,\si}^2\right)-{\hbar^2k_{so}^2\over
2m}
\end{eqnarray}
and the quantity $\xi_{l_{1,2}}(k_B)$ is given by \beq\lb{11}
\xi_{l_{1,2}}(k_B)= \sqrt{\xi^2+
\left[\varepsilon_l^-(k_B)\mp{\hbar\om_c\over2}
\left(1-2\lambda_{k,\si}^2\right)\right]^2}. \eeq The energy
$\varepsilon_l^-(k_B)$ is \beq \varepsilon_l^-(k_B)= {1\over
2}\left\{\varepsilon_l^{(0)}\left(k_B+k_{so}\right)-
\varepsilon_l^{(0)}\left(k_B-k_{so}\right)\right\} \eeq and $\xi$
takes the form \beq \xi = {\hbar^2k_{so}\over 2} \sqrt{{\om_c\over
m\hbar}\left(1-2\lambda_{k,\si}^2\right)}. \eeq We show that there
is
an hidden symmetry that encoded in the spectrum. Indeed, from the
Bloch theory it is easy to verify the relation \beq
\varepsilon_l^{(0)}(k_B)=\varepsilon_l^{(0)}(-k_B) \eeq which allows
us to show that the different energies are connected via
\begin{equation}
\varepsilon_{l,\eta=1}(k_B)=\varepsilon_{l,\eta=2}(-k_B),\qquad
\varepsilon_{l,\eta=3}(k_B)=\varepsilon_{l,\eta=4}(-k_B).
\end{equation}
The corresponding normalized eigenspinors $\te_{l,k_B,\eta}(n,\si)$
can be obtained by returning to the eigenvalue equation. This leads
to get
\begin{equation}\lb{9}
\te_{l,k_B,\eta=1,4}=N_{l,k_B,\eta=1,4}^{-{1\over
    2}}\tilde{\te}_{l,k_B,\eta=1,4}, \qquad
\te_{l,k_B,\eta=2,3}=N_{l,k_B,\eta=2,3}^{-{1\over
    2}}\tilde{\te}_{l,k_B,\eta=2,3}
\end{equation}
where the functions $\te_{l,k_B,\eta=1,4}$ take the form \beq
\tilde{\te}_{l,k_B,\eta=1,4}= \left( \begin{array}{c} {i\over
  \xi}\left[\varepsilon_l^-(k_B)-{\hbar\om_c\over2}
\left(1-2\lambda_{k,\si}^2\right)\mp\xi_{l_1}(k_B)\right]\\
0\\
0\\
1
\end{array}\right)
\eeq and the remaining quantities read as \beq
\tilde{\te}_{l,k_B,\eta=2,3}= \left( \begin{array}{c}
0\\
-{i\over \xi}\left[\varepsilon_l^-(k_B)+{\hbar\om_c\over2}
\left(1-2\lambda_{k,\si}^2\right)\pm\xi_{l_2}(k_B)\right]\\
1\\
0
\end{array}\right).
\eeq Note that, we have involved such notation \beq
{\te}_{l,k_B,\eta}\equiv \left( \begin{array}{c}
{\te}_{l,k_B,\eta}(n=0,\si=+1)\\
{\te}_{l,k_B,\eta}(n=0,\si=-1)\\
{\te}_{l,k_B,\eta}(n=1,\si=+1)\\
{\te}_{l,k_B,\eta}(n=1,\si=-1)
\end{array}\right)
\eeq in deriving the eigenspinors and the normalized constats can be
obtained as usual. Thus, it is not hard to show that there are \beq
N_{l,k_B,\eta}=\sum_{n=0}^1\sum_{\si=-1}^{+1}\left|\tilde{\te}_{l,k_B,\eta}(n,\si)\right|^2.
\eeq Finally, using the obtained spectrum it is easy to verify the
relation \beq\lb{12} N_{l,k_B,\eta=1,4}=N_{l,-k_B,\eta=2,3} \eeq
between different sub-bands entering in the game.

At this level, let us give a comment about the band structures.
Indeed, calculating the derivative of $\varepsilon_l^-(k_B)$ with
respect to $k_B$ at the zero point, we obtain the group velocity
\begin{equation}\label{gvelo}
{d\varepsilon_l^-(k_B) \over dk_B}|_{k_B=0} = v^{(0)}_l(k_{so})
\end{equation}
which is nothing but that corresponds to one-dimensional system.
Clearly, starting from the obtained spectrum one can observe that
the band splitting near $k_B=0$ is a linear function in terms of the
Bloch vector $k_B$. It is similar to what has experimentally been
observed for two-dimensional electron gazes~\cite{luo}. Our
conclusion is in agreement with what has been reached
in~\cite{smirnov}.

To close this section let us note that when the magnetic field is
switched off, we recover Bloch Hamiltonian amended by a free
particle living on the $z$-direction. To get the analysis
of~\cite{smirnov} simply we can add a harmonic potential in the same
direction. Our study shows that this potential can simply be
generated from the confinement due to the presence of $B$. On the
other hand, we will use the established tools to discuss the
symmetry in terms the polarizations.

\section{Polarization symmetries}

We start by noting that it has been shown that the system described
by the Hamiltonian~(\ref{ham}) is preserving the polarization
symmetries~\cite{smirnov}. Thus, it is natural to ask about such
symmetries for the present system in order to underline the
similarities and differences. To reply this question, we start
discussing the polarizations by adopting the definition
\beq\label{defpol} P_{l,\eta}^{(i)} (k_B)\equiv \langle
l,k_B,\eta|\si_i|l,k_B,\eta \rangle \eeq where $\si_i$ are the Pauli
spin operators, with $i=x,y,z$ and $\eta=1,2,3,4$. Therefore, we
need to evaluate different quantities entering in the game.

We start by analyzing each component separately. In doing so, we
write the identity operator in the $\{|l,k_B,n,\si \rangle\}$ basis
and take into account the structure of the obtained Bloch spinors.
This process leads to show that its expression in the $x$-direction
is given by \beq
P_{l,\eta}^{(x)}=\sum_{n=0}^1\sum_{\si',\si''=-1}^{+1}\te_{l,k_B,\eta}^*(n,\si')\
 \left(1-\delta_{\si',\si''}\right) \ \te_{l,k_B,\eta}(n,\si'')=0
\eeq as well as for  the $y$-direction \beq
P_{l,\eta}^{(y)}=\sum_{n=0}^1\sum_{\si',\si''=-1}^{+1}\te_{l,k_B,\eta}^*(n,\si')\
i^{\si''}\
  \left(1-\delta_{\si',\si''}\right) \te_{l,k_B,\eta}(n,\si'')=0
\eeq for all $l$ and $k_B$ in the Brillouin zone. Clearly, both of
polarizations longitudinal and perpendicular are nulls, which are in
agreement with the analysis of~\cite{smirnov}.

Let us now analyze the polarization
 along the $z$-direction. Indeed, for such case we show that (\ref{defpol})
reduces to the quantity \beq\lb{10}
P_{l,\eta}^{(z)}=\sum_{n=0}^1\sum_{\si=-1}^{+1}\te_{l,k_B,\eta}^*(n,\si)\
\si \ \te_{l,k_B,\eta}(n,\si).
\eeq To explicitly evaluate this form, we need to separate two cases
those governed by the four Bloch sub-bands and use of course the
obtained spinors. Thus, for $\eta=1,4$ we find \beq
P_{l,\eta}^{(z)}=N_{l,k_B,\eta=1,4}^{-1}\ \left\{{1\over
  \xi^2}\left[\varepsilon_l^-(k_B)-{\hbar\om_c\over2}\left(1-2\lambda_{k}^2\right)
\pm\xi_{l_1}(k_B)\right]^2-1\right\} \eeq and for $\eta=2,3$ we end
up with \beq P_{l,\eta}^{(z)}=N_{l,k_B,\eta=2,3}^{-1}\
\left\{1-{1\over
  \xi^2}\left[\varepsilon_l^-(k_B)+{\hbar\om_c\over2}
\left(1-2\lambda_{k}^2\right)\mp\xi_{l_2}(k_B)\right]^2\right\}.
\eeq This shows this component of polarizations has a finite value,
which makes difference with respect to those seen above. Moreover,
using (\ref{12}) and the relations \beq
\varepsilon_l^-(k_B)=-\varepsilon_l^-(-k_B), \qquad
\xi_{l_{1,2}}(k_B)\neq \xi_{l_{2,1}}(-k_B) \eeq
 we show that the symmetry relation for the polarizations verifies
\beq\lb{P} P_{l,\eta=1,4}^{(z)}(k_B) \neq
-P_{l,\eta=2,3}^{(z)}(-k_B). \eeq This tells us that the
polarization symmetries are not preserved for the present system.
These have been broken by the external magnetic field that interact
with the particles in terms of the gauge field. Its preservation can
be recovered by switching off $B$ and adding a confining potential
along $z$-direction~\cite{smirnov}. However, as we will in the next,
we can have such preservation of the symmetries by considering a
limit of strong $B$.

\section{Dynamical spin analysis }

In the beginning, let us note that the present system is sharing
some commun features with that describing the spin Hall effect
(SHE). In fact, to talk about SHE
 one has to consider the Rashba or Dressellhaus spin-orbit coupling as an
important ingredient in the Hamiltonian system. This suggests to
make a dynamical spin analysis of our system in order to underline
its behaviour. This can be done by determining the velocity and spin
current operators, which can be used to write down the corresponding
spin Hall conductivity $\si_{\sf H}^{\sf s}$.

To do our task we start by evaluating the velocity operators in
different directions. Indeed, from the Hamiltonian formalism one can
easily obtain the quantities
\begin{equation}\label{velo}
v_x= {\pi_x\over m}+{\hbar k_{so}\over m}\si_{z},\qquad v_z=
{\pi_z\over m}+{\hbar k_{so}\over m}\si_{x}
\end{equation}
which will be used to derive the spin current operator. To do this,
let us recall that due to the fact that spin is not a conserved
quantity in the presence of the spin-orbit coupling, there are
different definitions of such operator~\cite{ando}. But we adopt
that widely used where the spin current operator is defined as the
difference of the conserved densities for carriers with opposite
spins. This is
\begin{equation}\label{scop}
J_z^{\si_y}= {\hbar\over 4}(v_z\si_y + \si_y v_z)
\end{equation}
leading to the form
\begin{equation}\label{scop2}
J_z^{\si_y}= {\hbar \pi_z\over 2m}\si_y.
\end{equation}
Note that, $J_z^{\si_y}$ can be also obtained by using another
method. The advantage of this latter is to end up with an
appropriate form for $\si_{\sf H}^{\sf s}$, which we will do soon.

Reccal that SHE is a manifestation of a collective motion of the
spins of particles in an external electric field. This phenomenon is
characterized by $\si_{\sf H}^{\sf s}$ that is related to that for
the quantum Hall effect~\cite{zhang}. From linear response
formalism, it can be written in terms of the frequency $\om$ as
\cite{ando}
\begin{equation}\label{shcond}
\si_{\sf H}^{\sf S} (\om)=-{e\hbar \over iL^2}
\sum_{\al,\al'}{\langle \al| J_y^{\si_z} |\al'\rangle \langle
\al'|v_x |\al'\rangle \over  \eps_{\al} -\eps_{\al'} +\hbar\om +i0}\
{f(\eps_{\al})- f(\eps_{\al'}) \over \eps_{\al} -\eps_{\al'}}
\end{equation}
where we have set $|\al\rangle \equiv |l,k_B,n,\si\rangle$ and $
\eps_{\al} \equiv \varepsilon_{l,\eta}(k_B)$. It can be simplified
by introducing
 $J_y^{\si_z}$ in terms of the Heisenberg equation, namely
\begin{equation}\label{scop3}
J_z^{\si_y}= -{i\over 4} [H,\si_z].
\end{equation}
The corresponding matrix element in the basis $|\al\rangle$ can be
evaluated as
\begin{equation}
\langle \al|J_z^{\si_y} |\al'\rangle= -{i\over 4} (\eps_{\al}
-\eps_{\al'}) \langle \al| \si_z |\al'\rangle.
\end{equation}
Injecting this in (\ref{shcond}), we end up with
\begin{equation}\label{shcond2}
\si_{\sf H}^{\sf S} (\om)= {e\hbar \over 4L^2} \sum_{\al,\al'}{
\langle \al| \si_z |\al'\rangle
 \langle \al'|v_x |\al'\rangle
\over  \eps_{\al} -\eps_{\al'} +\hbar\om +i0}\ f(\eps_{\al})-
f(\eps_{\al'}).
\end{equation}
To explicitly determine such conductivity one may use the
self-consistent Born approximation in similar way to~\cite{ando}. On
the other hand, (\ref{shcond2}) can be analyzed by adopting the high
and low frequencies regimes to reach different conclusions.

\section{Analysis of two cases}

We are wondering to analyze different limits in order to
characterize the system behaviour. From this section we start by
considering the cases where the Rashba coupling constant and the
periodic potential are nulls. These will allow us to end up with a
Landau problem amended either with a periodic potential or the
Rashba term in the presence of the magnetic field.

\subsection{Without Rashba coupling}

 The first one
concerns the Rashba coupling constant when is absent. In fact, we
return back to the former studies and show its restrictions to such
case. In doing so, we derive the corresponding spectrum as well as
discuss the polarizations symmetries. Before doing our job, let us
fix the Hamiltonian that describes the present limit. This can
easily be obtained from (\ref{totham}) to get
\begin{equation}\label{ham0kso}
H|_{k_{so}=0}= {1\over 2m}\left[p_z^2 + \left(p_x-{eB\over c}z
\right)^2 \right] +U(x)
\end{equation}
which is nothing but the Landau problem written in the Landau gage
and submitted to the constraint of a periodic potential. Clearly,
without $B$ we have a Bloch theory along $x$-direction plus a wave
plane along $z$-direction.

To get the eigenvalues and the eigenstates, we algebraically
diagonalize the above Hamiltonian. To proceed, let us define the
annihilation and creation operators as
\begin{equation}\label{ham0kso2}
c={l_B\over\sq2\hbar}\left[p_z+i\left(p_x-{eB\over
c}z\right)\right], \qquad
c^{\da}={l_B\over\sq2\hbar}\left[p_z+i\left(p_x-{eB\over
  c}z\right)\right]
\end{equation}
where $l_B=\sq{{c\hbar\over eB}}$ is the magnetic length. It is easy
to verify the commutation relation
\begin{equation}
\left[c, c^{\da}  \right] = \mathbb{I}.
\end{equation}
In terms of  these operators, (\ref{ham0kso}) can be written as
\begin{equation}
H|_{k_{so}=0}= {\hbar\om_c} \left(c^{\da} c+ {1\over 2}\right) +
U(x).
\end{equation}
Therefore the corresponding spectrum is given by
\begin{equation}
E_n|_{k_{so}=0} (k_B) ={\hbar\om_c}  \left(n+ {1\over 2}\right),
\qquad |n,k_B\rangle, \qquad n=0,1,2, \cdots
\end{equation}
where the states explicitly read as
\begin{equation}
|n\rangle={(a^{\da})^{n}\over \sqrt{n!}} |0\rangle, \qquad \langle
x|k_B\rangle={1\over\sq L_0}e^{ik_Bx}u_{k_B}(x).
\end{equation}
To close this part, we note that the eigenvalues depend only on $n$
and are completely independent of the Bloch momentum $\hbar k_B$.
Due to the lack of dependence of the energy on $k_B$, the degeneracy
of each level is enormous.

\subsection{Without periodic potential}

Let us study another case that corresponds to $U(x)$ equal to zero.
This will allows us to characterize  the nature of the present
system when $U(x)$ is switched off. In doing so, we give the
corresponding Hamiltonian, which can be deduced from (\ref{totham})
as \beq\lb{u=0} H|_{U(x)=0}={1\over2m}\left(\pi_z^2+\pi_x^2\right)-
{\hbar^2 k_{so}\over m}\left(\si_x\pi_z-\si_z\pi_x\right). \eeq
Before proceeding we note that an analogue operator has been
considered in different contexts but when only the $x$ an $y$
coordinates are involved, for instance one may
see~\cite{schliemann03a}. Here is different because of the last term
which is including diagonal and off-diagonal matrices.

To get the spectrum of (\ref{u=0}) we adopt an analysis similar to
that has been reported in subsection (4.1) except that we take
 $U(x)= 0$.
We also note that for each state we can still choose $ k_x$ freely,
thus we have degenerate energy levels. If we assume our space is
finite, for definiteness we choose a rectangle of dimensions $ L_{x}
\times L_{y}$, the number of states in a level is finite and can be
calculated. The eigenvalues of $ p_{x}$ are quantized as $
k_x=\frac{2\pi}{L_{y}}j$ where $j$ is a quantum number. Therefore,
after diagonalizing  (\ref{u=0})
we end up with the relations
\begin{equation}\lb{7u=0}
\varepsilon_{l,\eta=1,2}(j)=\varepsilon_l^+(j)-\xi_{l_{1,2}}(j),\qquad
\varepsilon_{l,\eta=3,4}(j)=\varepsilon_l^+(j)+\xi_{l_{2,1}}(j)
\end{equation}
where $\varepsilon_l^+(j)$ takes the form
\begin{eqnarray}
\varepsilon_l^+(j) = {\hbar^2\over 2m}\left({2\pi j\over
L_y}\right)^2+ \hbar\om_c\left(1-2\lambda_{j,\si}^2\right)
\end{eqnarray}
and  $\xi_{l_{1,2}}(j)$ reads as \beq\lb{11u=0} \xi_{l_{1,2}}(j)=
\sqrt{\xi^2+ \left[\varepsilon_l^-(j)\mp{\hbar\om_c\over2}
\left(1-2\lambda_{j,\si}^2\right)\right]^2}. \eeq The energy
$\varepsilon_l^-(j)$ is given by
\beq
\varepsilon_l^-(j)={\hbar^2\over m} \left({2\pi j\over
L_y}\right)k_{so} \eeq
and the quantity $\xi$ is \beq \xi =
{\hbar^2k_{so}\over 2} \sqrt{{\om_c\over
m\hbar}\left(1-2\lambda_{j,\si}^2\right)} \eeq where the parameter
$\lambda_{j,\si}$ becomes \beq
\lambda_{j,\si}=-\sq{{\hbar\over2m\om_c}}\left({2\pi j\over L_y}+\si
k_{so}\right). \eeq These lead to derive the normalized eigenspinors
$\te_{l,j,\eta}(n,\si)$ as
\begin{equation}\lb{9u=0}
\te_{l,j,\eta=1,4}=N_{l,j,\eta=1,4}^{-{1\over
    2}}\tilde{\te}_{l,j,\eta=1,4}, \qquad
\te_{l,k_B,\eta=2,3}=N_{l,j,\eta=2,3}^{-{1\over
    2}}\tilde{\te}_{l,j,\eta=2,3}
\end{equation}
where the spinors $\te_{l,j,\eta=1,4}$ read as \beq
\tilde{\te}_{l,j,\eta=1,4}= \left( \begin{array}{c} {i\over
  \xi}\left[\varepsilon_l^-(j)-{\hbar\om_c\over2}
\left(1-2\lambda_{j,\si}^2\right)\mp\xi_{l_1}(j)\right]\\
0\\
0\\
1
\end{array}\right)
\eeq and $\te_{l,j,\eta=2,3}$ are given by \beq
\tilde{\te}_{l,j,\eta=2,3}= \left( \begin{array}{c}
0\\
-{i\over \xi}\left[\varepsilon_l^-(j)+{\hbar\om_c\over2}
\left(1-2\lambda_{j,\si}^2\right)\pm\xi_{l_2}(j)\right]\\
1\\
0
\end{array}\right).
\eeq

It is natural ask about the polarization symmetries in the present
case. The answer can be obtained by evaluating different components
and to obtain for $\eta=1,4$ the result \beq
P_{l,\eta}^{(z)}=N_{l,j,\eta=1,4}^{-1}\ \left\{{1\over
  \xi^2}\left[\varepsilon_l^-(j)-{\hbar\om_c\over2}\left(1-2\lambda_{j,\si}^2\right)
\pm\xi_{l_1}(j)\right]^2-1\right\} \eeq and for $\eta=2,3$ we have
\beq P_{l,\eta}^{(z)}=N_{l,j,\eta=2,3}^{-1}\ \left\{1-{1\over
  \xi^2}\left[\varepsilon_l^-(j)+{\hbar\om_c\over2}
\left(1-2\lambda_{j,\si}^2\right)\mp\xi_{l_2}(j)\right]^2\right\}.
\eeq By showing the relation
\begin{equation}
P_{l,\eta=1,4}^{(z)}(j)|_{U(x)=0} \neq
-P_{l,\eta=2,3}^{(z)}(-j)|_{U(x)=0}.
\end{equation}
we conclude that such symmetries  are also not preserved in the
present case.  This close the analysis of the Rashba and periodic
potential terms. In the next, we will study to other cases.

\section{Strong magnetic field limit}

It is interesting to analyze the case where the champ magnetic is
strong. Such interest comes from the fact that this limit is
important for the quantum Hall effect. Thus, it will be a good task
to deal with such limit and derive different results.

Recall that $B$ is included in the Hamiltonian for the
$z$-direction. Requiring that $B$ is strong,  we can approximate
(\ref{hamzd}) as
\begin{equation}
H^{(1)}_z|_{\rm sB}= {\hbar^2k_z^2\over 2m}+{1\over
2m}\left({eB\over
  c}\right)^2z^2
\end{equation}
which is nothing but the one-dimensional harmonic oscillator of
frequency $\om_c$, with the abbreviation $({\rm sB})$ means strong
magnetic field. Therefore, the corresponding spectrum is given by
\begin{equation}
E_n|_{\rm sB}= \hbar \om_c\left(n+{1\over 2}\right).
\end{equation}
This modifies the $H^{(1)}$ eigenvalues as
\begin{equation}
\varepsilon_{l,n,\si}^{(1)}(k_B)|_{\rm sB}=
\varepsilon_{l}^{(0)}\left(k_B+\si k_{so}\right)-
{\hbar^2k_{so}^2\over 2m}+\hbar\om_{c} \left(n+{1\over 2}\right)
\end{equation}
as well as the matrix element
\begin{equation}
\langle n|k_z|n'\rangle=\pm i\delta_{n,n'\pm1}\sqrt{{m\om_c\over
2\hbar} {\left(n+{1\over 2}\mp{1\over 2}\right)}}.
\end{equation}
Note that, what makes difference with respect to the previous
analysis is that the parameter $\lam_{k,\si}$ is absent. This of
course will offer different simplifications and determine
interesting results concerning the polarizations.

To be coherent with our analysis, we consider the two first
transverses modes and derive the whole spectrum. Doing this to get
\begin{equation}\lb{sB}
\varepsilon_{l,\eta=1,2}(k_B)|_{\sf
sB}=\varepsilon_l^+(k_B)-\xi_{l_{1,2}}(k_B)|_{\sf sB},\qquad
\varepsilon_{l,\eta=3,4}(k_B)|_{\sf
sB}=\varepsilon_l^+(k_B)+\xi_{l_{2,1}}(k_B)|_{\sf sB}
\end{equation}
where different quantities changed to \beq
\varepsilon_l^+(k_B)|_{\rm sB}= {1\over 2}\left\{
\varepsilon_l^{(0)}\left(k_B+k_{so}\right)+
\varepsilon_l^{(0)}\left(k_B-k_{so}\right) \right\}
 +\hbar\om_c-{\hbar^2k_{so}^2\over 2m}
\eeq as well as \beq \xi_{l_{1,2}}(k_B)|_{\rm sB}= \sqrt{\xi^2|{\rm
sB}+\left(\varepsilon_l^-(k_B) \mp{\hbar\om_c\over2}\right)^2} \eeq
and we have set \beq \xi|_{\rm sB} = {\hbar^2k_{so}\over
m}\sqrt{{m\om_c\over 2\hbar}}. \eeq The corresponding Bloch spinors
are given by
\begin{equation}
\tilde{\te}_{l,k_B,\eta=1,4}|_{\rm sB}= \left( \begin{array}{c}
{i\over \xi|_{\rm sB}} \left[\varepsilon_l^-(k_B)-{\hbar\om_c\over2}
\mp\xi_{l_1}(k_B)|_{\rm sB}\right]\\
0\\
0\\
1
\end{array}\right)
\end{equation}
and the other sub-band read as
\begin{equation}
\tilde{\te}_{l,k_B,\eta=2,3}|_{\rm sB}= \left( \begin{array}{c}
0\\
-{i\over \xi|_{\rm sB}}\left[\varepsilon_l^-(k_B)+{\hbar\om_c\over2}
\pm\xi_{l_2}(k_B)|_{\rm sB}\right]\\
1\\
0
\end{array}\right).
\end{equation}

Is relevant to discuss the polarization symmetries in the present
case to underline what makes differences with the previous analysis.
On the other hand, the above analysis is sharing some commun
features with that realized in~\cite{smirnov}. Indeed, using the
above tools to show that for $\eta=1,4$ we have \beq
P_{l,\eta}^{(z)}|_{\rm
sB}=N_{l,k_B,\eta=1,4}^{-1}\times\left\{{1\over
  \xi^2|_{\rm sB}} \left[\varepsilon_l^-(k_B)-{\hbar\om_c\over2}
\mp\xi_{l_1}(k_B)|_{\rm sB} \right]^2-1\right\} \eeq as well as for
$\eta=2,3$ \beq P_{l,\eta}^{(z)}|_{\rm
sB}=N_{l,k_B,\eta=2,3}^{-1}\times\left\{1-{1\over
  \xi^2|_{\rm sB}} \left[\varepsilon_l^-(k_B)+
{\hbar\om_c\over2}\pm\xi_{l_2}(k_B)|_{\rm sB} \right]^2\right\} \eeq
Finally, we end up with
\begin{equation}
P_{l,\eta=1,4}^{(z)}(k_B)|_{\rm sB} =
-P_{l,\eta=2,3}^{(z)}(-k_B)|_{\rm sB}.
\end{equation}
We reach the conclusion that the polarizations for strong magnetic
field are conserved. This is an interesting results because it gives
another way to talk about such symmetries, which is different from
what has been developed in~\cite{smirnov}. In fact, these
polarizations is actually controlled by the external parameter $B$
and can be adjusted by switching on $B$ to diffrent values.

\section{Critical point analysis}

As we have noticed before in diagonalizing the Hamiltonian of the
present system there is a critical point, i.e. $\lam_{k,\si}=\pm
\sq{{1\over 2}}$, with $\si=\pm 1$. It is a good task to study the
effect of this point on the obtained results and underline what it
will be new. To do this, we return to the Hamiltonian describing a
system like harmonic oscillator where $\lam_{k,\si}$ is included and
resort the corresponding spectrum to see what makes differences with
respect to the standard case.

Taking into account such critical point, we can establish a relation
between the Bloch wave vector and the Rashba one. More precisely, we
have \beq k_{so}|_{\si=1} = \mp \sqrt{{m\om_c\over 2\hbar}} - k_B,
\qquad k_{so}|_{\si=-1} = \pm \sqrt{{m\om_c\over 2\hbar}} + k_B.
\eeq It is convenient to write these relations as functions of the
magnetic length $l_B$, such as \beq k_{so}|_{\si=1} = \mp {1\over
\sqrt{2}l_B} - k_B, \qquad k_{so}|_{\si=-1} = \pm {1\over
\sqrt{2}l_B} + k_B. \eeq These relations are interesting in sense
that one vector can be determined in terms of the other. Moreover,
they are $B$-dependent and clearly for a strong $B$ one of them can
be ignored where the remaining vector will be controlled by $B$. In
this limit for instance the area of the Hall droplet can be written
as \beq S= 2\pi l_B^2 \equiv {\pi\over
\left(k_{so}|_{\si=1}\right)^2}. \eeq

Now let us return to write the corresponding
 Hamiltonian, which is
\begin{equation}\label{h1o2}
H^{(1)}_z|_{\pm\sq{{1\over 2}}}=\hbar\om_c a^{\dagger}a.
\end{equation}
This leads to end up  with the spectrum
\begin{equation}\label{cgra}
\varepsilon_{l,n,\si}^{(1)}(k_B)|_{z,\pm \sq{{1\over 2}}}=\hbar\om_c
n, |n\rangle = {(a^{\da})^n\over \sqrt{n!}}|0\rangle, \qquad n=0,1,2
\cdots.
\end{equation}
The first remark one can underline is that actually the Hilbert
space of such spectrum has an isolated point that correspond to a
zero mode energy. This is not a suprising result because there are
some systems who are behaving like this. In fact, electrons in
graphene, which behave like massless Dirac fermions (Majorana
fermions), in the presence of a magnetic field have such kind of
zero mode energy~\cite{jellal}. More precisely, for one single
particle in graphene can be described by the Dirac operator
\begin{equation}\label{hadm}
H_{\sf D}={i\over \sqrt{2}} v_{\sf F}\left(%
\begin{array}{cc}
  0 & d^{\dag} \\
  -d & 0 \\
\end{array}%
\right)
\end{equation}
where $v_{\sf F}\thickapprox\frac{c}{100}$ is the Fermi velocity,
which will be set to one,  and the many-body effects are neglected.
The annihilation and creation operators, in the complex notation
$z=x+iy$, are given by
\begin{equation}\label{ocra}
d= 2{\pa\over \partial{z}}+\frac{B}{2}\bar{z}, \qquad
    d^{\dag}=-2{\pa \over \partial{\bar{z}}}+\frac{B}{2}z.
\end{equation}
They verify the commutation relation
\begin{equation}\label{recom}
    [d, d^{\dag}]=2B.
\end{equation}
Now let us look at its square, which is
\begin{equation}\label{hadsq}
    H_{\sf D}^{2}= {1\over 2}\left(%
\begin{array}{cc}
 d^{\dag}  d& 0 \\
  0 &  d d^{\dag}
\end{array}%
\right).
\end{equation}
Clearly, the first matrix element is similar to that of
(\ref{h1o2}). The corresponding
 wavefunctions $\Psi$ should be
written in an appropriate form. This is
\begin{equation}\label{eigvec}
 \Psi_{m,n}=\left(%
\begin{array}{c}
 \psi_{m,n} \\
\psi_{m-1,n}
\end{array}%
\right)
\end{equation}
where the eigenfunctions $\psi_{m,n}$ are given by
\begin{equation}\label{eigfun}
    \psi_{m,n} (z,\bar z)=\frac{(-1)^{m}\sqrt{B^{m}m!}}{\sqrt{2^{n+1}
\pi(m+n)!}}z^{n}L^{n}_{m}\left(\frac{z\bar{z}}{2}\right)
e^{-\frac{B}{4}z\bar{z}},\qquad
    m,n=0,1,2\cdots.
\end{equation}
Their Landau levels take the form
\begin{equation}\label{landlev}
    E_{{\sf D}}^{2}(m)=Bm.
\end{equation}
Finally, we end up with the zero mode energy for such system, namely
\begin{equation}\label{eigfunf}
    \Psi^{(0,n)}=\left(%
\begin{array}{c}
  \psi_{0,n}\\
0
\end{array}%
\right)\cdot
\end{equation}
This is showing the similarity with the previous case that
corresponds to the Hamiltonian $H^{(1)}_z$ at the critical point.

Now let us return to derive the whole spectrum in the present case.
Indeed, as an immediate consequence the eigenvalues of $H^{(1)}$
becomes
\begin{equation}
\varepsilon_{l,n,\si}^{(1)}(k_B)|_{\pm \sq{{1\over 2}}}=
\varepsilon_{l}^{(0)}\left(k_B+\si k_{so}\right)-
{\hbar^2k_{so}^2\over 2m}+\hbar\om_{c} n.
\end{equation}
The matrix element (\ref{mtelt}) now reads as
\begin{equation}
\langle n|k_z|n'\rangle=\pm i\delta_{n,n'\pm1}\sqrt{{m\om_c\over
2\hbar} {\left(n+{1\over 2}\mp{1\over 2}\right)}}.
\end{equation}

Focussing on the two first transverse modes, we show that different
quantities forming the total spectrum take other forms. Indeed, we
find
\begin{eqnarray}
\varepsilon_l^+(k_B)|_{\pm\sq{{1\over2}}} = {1\over 2} \left\{
\varepsilon_l^{(0)}\left(k_B+k_{so}\right)+
\varepsilon_l^{(0)}\left(k_B-k_{so}\right)\right\}+{\hbar\om_c\over2}-{\hbar^2k_{so}^2\over
2m}
\end{eqnarray}
as well as \beq \xi_{l}(k_B)|_{\pm\sq{{1\over2}}}
=\sq{\xi^2|_{\pm\sq{{1\over2}}}+
\left(\varepsilon_l^-(k_B)+{\hbar\om_c\over2}\right)^2} \eeq
where $\xi$ at the critical point is given by \beq
\xi|_{\pm\sq{{1\over2}}}  ={\hbar^2k_{so}\over
m}\sq{{m\om_c\over2\hbar}}. \eeq These leading to end up with the
energies for different sub-bands, namely \beq
\varepsilon_{l,\eta=1,2}(k_B)=\varepsilon_l^+(k_B)-\xi_{l}(k_B)|_{\pm\sq{{1\over2}}},
\qquad
\varepsilon_{l,\eta=3,4}(k_B)=\varepsilon_l^+(k_B)+\xi_{l}(k_B)|_{\pm\sq{{1\over2}}}.
\eeq The corresponding Bloch spinors read as
\begin{equation}
\tilde{\te}_{l,k_B,\eta=1,4}|_{\pm\sq{{1\over2}}}= \left(
\begin{array}{c} {i\over \xi|_{\pm\sq{{1\over2}}}
}\left[\varepsilon_l^-(k_B)+{\hbar\om_c\over2}
+\xi_l(k_B)|_{\pm\sq{{1\over2}}}\right]\\
0\\
0\\
1
\end{array}\right)
\end{equation}
as well as for other modes
\begin{equation}
\tilde{\te}_{l,k_B,\eta=2,3}|_{\pm\sq{{1\over2}}}= \left(
\begin{array}{c}
0\\
-{i\over
\xi|_{\pm\sq{{1\over2}}}}\left[\varepsilon_l^-(k_B)+{\hbar\om_c\over2}
-\xi_l(k_B)|_{\pm\sq{{1\over2}}}\right]\\
1\\
0
\end{array}\right).
\end{equation}

The above materials can be used to analyze the polarization
symmetries at the critical point. In fact, we show that for
longitudinal and perpendicular cases the polarizations are zero.
However it is finite along $z$-direction and this can be traduced by
the obtained results. Indeed,
 for $\eta=1,\cdots,4$ we obtain
\beq
P_{l,\eta}^{(z)}|_{\pm\sq{{1\over2}}}=N_{l,k_B,\eta=1,4}^{-1}|_{\pm\sq{{1\over2}}}
\times\left\{{1\over
  \xi^2|_{\pm\sq{{1\over2}}}} \left[\varepsilon_l^-(k_B)+{\hbar\om_c\over2}
+\xi_{l}(k_B)|_{\pm\sq{{1\over2}}} \right]^2-1\right\} \eeq and for
$\eta=2,3$ we have \beq
P_{l,\eta}^{(z)}|_{\pm\sq{{1\over2}}}=N_{l,k_B,\eta=2,3}^{-1}|_{\pm\sq{{1\over2}}}
\times\left\{1-{1\over
  \xi^2|_{\pm\sq{{1\over2}}}} \left[\varepsilon_l^-(k_B)+
{\hbar\om_c\over2}-\xi_{l}(k_B)|_{\pm\sq{{1\over2}}}
\right]^2\right\}. \eeq These allow us to end up with the relation
\begin{equation}
P_{l,\eta=1,4}^{(z)}(k_B)|_{\pm\sq{{1\over2}}} \neq
-P_{l,\eta=2,3}^{(z)}(-k_B)|_{\pm\sq{{1\over2}}}
\end{equation}
which means that there is no preservation of such symmetries as we
have seen in the standard case. With this we end the analysis of
four cases and we will see the last one.

\section{Periodic structure without magnetic field}

To complete the present analysis it is important to treat the case
of the periodic structure without magnetic field. This will allow us
to get new results and make contact with other proposals. In doing
so, we simply return to the total Hamiltonian and derive the
corresponding spectrum when $B$ is zero.

To start let us consider the Hamiltonian describing the system in
the absence of $B$. This can be obtained from (\ref{1}) to end up
with
\begin{equation}
\label{hamb=0} H|_{B=0}={{\hbar}^2{\vec{k}}^2\over2m}-{{\hbar}^2
k_{so}\over m} \left(\si_{x}k_z-\si_{z}k_x\right)+U(x).
\end{equation}
Clearly for arbitrary periodic potential $U(x)$, the eigenvalues
(\ref{13}) and eigenstates (\ref{13p}) reduce to those of wave
planes. Indeed, it is easy to notice \beq |n\rangle \lga
|k_z\rangle, \qquad \varepsilon_{l,n,\si}^{(1)}(k_B)|_z \lga
{\hbar^2k_z^2\over 2m}. \eeq These can be used to show that the
matrix element takes the form \beq \langle k_z|k_z|k_z'\rangle=
i\delta_{k_z,k_z'}k_z. \eeq Combining all together to get the
energies corresponding to different sub-bands. Note that, we have
now only two possible values for $\eta$, i.e. (1,2). Thus, we have
\begin{eqnarray}\lb{si}
\varepsilon_{l,\eta=1,2}(k_B,k_z)&=& {1\over
2}\left[{\varepsilon_l^{(0)}\left(k_B+k_{so}\right)
+\varepsilon_l^{(0)}\left(k_B-k_{so}\right)}\right]+{\hbar^2k_z^2\over
2m}
-{\hbar^2k_{so}^2\over 2m} \pm\nonumber\\
&&
\pm\sqrt{\bigg(\varepsilon_l^-(k_B)\bigg)^2+\left({\hbar^2k_{so}k_z\over
 m}\right)^2}.
\end{eqnarray}

At this point, there are two limits those should be discussed to
characterize the present situation. The first one is concerning the
case where we assume that  $k_B=0$ and $k_z>0$. Returning to the
spectrum to obtain \beq\lb{ff}
\varepsilon_{l,\eta=1,2}^{2D}(k_B=0,k_z)=\varepsilon_l^{(0)}\left(k_{so}\right)+{\hbar^2k_z^2\over
2m}-{\hbar^2k_{so}^2\over
 2m}\pm{\hbar^2k_{so}k_z\over
 m}
\eeq This tell us that the energy branch with $\eta=2$ has a
 minimum at $k_z=k_{so}$ for all band $l$. These analytical results are in contrast
to what was numerically predicted in~\cite{demi}.

The second limit can be fixed by requiring that $U(x)= 0$. In this
case, one can show that the corresponding energies are
\begin{equation}\label{rash}
\varepsilon_{\si,\eta=1,2}(k_x,k_z) = {\hbar^2\over 2m}
\left(k_x^2+k_z^2 +2\si k_{so}\sqrt{k_x^2 +k_z^2} \right).
\end{equation}
In conclusion (\ref{rash}) is in agreement with the Rashba energy
for the present system without periodic potential and magnetic
field.

\section{Conclusion}

We have investigated the basic features of two-dimensional periodic
structures with Rashba spin-orbit interaction in the presence of  an
external magnetic field $B$. This latter allowed us to end up with a
confining potential along $z$-direction that has been used to deal
with different issues. In particular, it has been served to split
the corresponding Hamiltonian into two parts and one of them is
obtained to behave like one-dimensional harmonic oscillator. This
decomposition was usefull in  sense that different spectrum's are
obtained and leading to the total one. To make comparisons with
other proposals, we have restricted ourselves to the two first
transverse modes. These are used to derive the eigenvalues and the
Bloch spinors for four sub-bands $\eta=1,\cdots ,4$. Looking at the
obtained spectrum we have shown that there is an hidden symmetry,
which has been used to talk about the polarizations.

After getting the eigenvalues and the corresponding eigenspinors, we
have analyzed the polarization symmetries by evaluating their
components in different directions. As consequence, the longitudinal
and perpendicular are obtained to be nulls except that along
$z$-direction where a finite values is found. Looking at the
symmetry with respect to the Bloch vector, we have shown that the
presence of the magnetic field breaks such symmetries in
disagreement with the results obtained in~\cite{smirnov}. As another
task, we have investigated the dynamical spin by determining some
relevant quantities those used to write down the appropriate spin
Hall conductivity.

Subsequentely, different limits have been studied. Indeed, we have
analyzed the case when the Rashba coupling is null and this allowed
us to end up with a Landau problem submitted to the
 periodicity constraint. Therefore, the spectrum is easily obtained
by noting that the corresponding energies are degenerates and their
degree of degeneracy can be determined in terms of the Bloch vector.
Moreover, we have considered another case that is $U(x)=0$, which
lead to get the Landau problem amended with the Rashba coupling in
the presence of the magnetic field. Solving the eigenvalue equation
we have derived the spectrum that has been used to conclude that the
polarizations are not preserved.

Inspecting other limits, the strong magnetic field case has been
considered. This allowed us to get different conclusions and derive
interesting results. Indeed, after getting the corresponding
spectrum, we have tackled the symmetries problem. More precisely, we
have shown that in the present situation the polarizations are
conserved and therefore offered another way to talk about such
symmetries. More importantly, such analysis allowed us to make
contact with the proposal developed in~\cite{smirnov} and in
particular to show how one can recover the symmetry preservation's
from the present work. Moreover, the obtained polarization's are
magnetic field dependent and therefore it can be adjusted by
switching such field to any value.

On the other hand, we have studied in details one important issue
that allowed us to make contact with other systems. In fact, the
critical point $\lam_{k,\si}=\pm \sq{1\over 2}$ that appeared as a
singularity in diagonalizing the Hamiltonian $H^{(1)}_z$ has been
analyzed. Resorting the corresponding spectrum we have ended up with
a zero mode energy in the Hilbert space of $H^{(1)}_z$. In fact, a
comparison to the Dirac fermions in the presence of the magnetic
field is established.

Finally, to get more information about the present system we have
investigated the case where $B$ is switched off. This has been done
by deriving the spectrum for $\eta=1,2$, which are used to study two
limits. Indeed, assuming that $k_B=0$ where $k_z>0$, we have shown
that the energy for $\eta=2$ has a
 minimum at $k_z=k_{so}$ for all band $l$, which is
in agreement with that has been numerically predicted
in~\cite{demi}. Moreover, requiring that $U(x)=0$, we have recovered
the Rashba energy.

\section*{Acknowledgment}

This work was completed during a visit of AJ to the Abdus Salam
Centre for Theoretical Physics (Trieste, Italy) in the framework of
junior associate scheme. He would like to acknowledge the financial
support of the centre.

\end{document}